\RequirePackage[fleqn]{amsmath}
\documentclass[12pt]{iopart}
\usepackage{bm}
\usepackage{cite}
\usepackage{graphicx}
\begin{document}
\title[Parrondo Game with Short-Term Memory]{Winning in Sequential Parrondo Games by Players with Short-Term Memory}
\author{K W Cheung$^1$, H F Ma$^1$, D Wu$^1$, G C Lui$^1$ and K Y Szeto$^{1,2}$}
\address{$^1$Department of Physics, The Hong Kong University of Science and Technology, Clear Water Bay, Hong Kong, HKSAR, China}
\ead{${^2}$phszeto@ust.hk}
\begin{abstract}
The original Parrondo game, denoted as AB3, contains two independent games: A and B. The winning or losing of A and B game is defined by the change of one unit of capital.  Game A is a losing game if played continuously, with winning probability $p=0.5-\epsilon$, where $\epsilon=0.003$. Game B is also losing and it has two coins: a good coin with winning probability $p_g=0.75-\epsilon$ is used if the player`s capital is not divisible by $3$, otherwise a bad coin with winning probability $p_b=0.1-\epsilon$ is used. 
Parrondo paradox refers to the situation that the mixture of A and B game in a sequence leads to winning in the long run. The paradox can be resolved using Markov chain analysis. We extend this setting of Parrondo game to involve players with one-step memory. The player can win by switching his choice of A or B game in a Parrondo game sequence.  If the player knows the identity of the game he plays and the state of his capital, then the player can win maximally. 
On the other hand, if the player does not know the nature of the game, then he is playing a (C,D) game, where either (C=A, D=B), or (C=B,D=A). For player with one-step memory playing the AB3 game, he can achieve the highest expected gain with switching probability equal to $3/4$ in the (C,D) game sequence. This result has been found first numerically and then proven analytically. 
Generalization to AB mod($M$) Parrondo game for other integer $M$ has been made for the general domain of parameters 
$p_b<p=0.5=p_A <p_g$. We find that for odd $M$, Parrondo effect does exist. However, for even $M$, there is no Parrondo effect for two cases: initial game is A and initial capital is even, or initial game is B and initial capital is odd. There is still a possibility of Parrondo effect for the other two cases when $M$ is even: initial game is A and initial capital is odd, or initial game is B and initial capital is even. These observations from numerical experiments can be understood as the factorization of the Markov chains into two distinct cycles.
Discussion of these effects on games is also made in the context of feedback control of the Brownian ratchet. 
\end{abstract}
\submitto{\JSTAT}
\maketitle
\section{Introduction}
The concept of Brownian Ratchets, proposed by Smoluchowsky~\cite{smoluchowski_experimentell_1927} about a century ago, considers non-equilibrium devices where fluctuations can be rectified to produce directional transport of particles along a periodic asymmetric potential. Later, Feynman, in his famous lecture on the ratchet-and-pawl setup, discussed the Brownian motor, which is a Brownian ratchet with a load~\cite{feynman_feynman_1963,parrondo_energetics_2002}. About 20 years ago, the flashing ratchet with scheduled on/off switching of an asymmetric potential was discussed~\cite{ajdari_mouvement_1992,astumian_fluctuation_1994,hanggi_brownian_1996,reimann_brownian_2002} and inspired the invention of the Parrondo game~\cite{parrondo_how_1996}, which has a paradoxical scenario that two losing games can be combined to become a winning game. It then generates many works on different switching strategies for the Parrondo Game to maximize the gain. Dinis has found an optimal periodic sequence that maximizes the expected payoff~\cite{dinis_optimal_2008} and this work has been generalized using Genetic Algorithm for much longer sequence by Wu and Szeto~\cite{wu_applications_2014}. Chaotic game sequences have also been investigated by Tang et al.~\cite{tang_investigation_2004}. Apart from the original Parrondo game, other types of Parrondo games have been constructed such as the Quantum Parrondo Game~\cite{flitney_quantum_2002}, history-dependent Parrondo Game~\cite{parrondo_new_2000} and Extended Parrondo Game that incorporate different B games without the diffusive A game~\cite{wu_extended_2014}. Here we address the benefits for player with finite memory in various switching strategies when playing the original Parrondo Game. We find that the information accessible by the player is critical to his success in winning. Information obtained by communication or memory provides a feedback to the player in his selection of strategy for the next game to improve his chance of winning. This is an example of the feedback-control algorithm in control theory for improving the overall performance.  For players with one-step memory, we find that certain switching strategies can lead to optimal winning. In this paper, we will first outline the solution of the original Parrando game. We then illustrate the importance of various types and the amount of information accessible to the player in enhancing his chance of winning through a protocol of ``feedback-control''.
\par
In layman term, the paper addresses the problem of optimizing capital gain for two players, Edward and Frank, who enter a casino and play on two slot machines called ``Charles'' and ``David''. Both players are poor in counting, so they lose count of the amount of capital they have. However, they have one-step memory in the sense that they remember which slot machine they used in the last game. Edward gets some favours from the casino boss and knows that ``Charles'' is equipped with the software corresponding to the A game in the original Parrondo paper, and ``David'' is equipped with the software corresponding to the B game. As for Frank, he does not know this information, though he is assured that one of the slot machines is installed with A game and the other is with B game software. Thus, Frank has less information than Edward in playing the game. They both consult a physicist about how to play in these two slot machines so that they can exploit the Parrondo effect to win in the long run, knowing that if they continuously play on one machine they will lose. The physicist solves this problem for both Edward and Frank and provides an algorithm for switching so that they will win. From now on, the game sequence for Edward is labelled by (A, B) and is called the AB game, because he knows the software installed in the two slot machines. The game sequence for Frank is labelled by (C,D) and is called the CD game, for he does not know the software installed. 
\section{Parrondo game with random switching}
The original Parrondo game contains two independent games: A and B. Game A is a coin-tossing game with a winning probability $p=0.5-\epsilon \, \left(\epsilon>0 \right)$, so that A is a losing game in the long run. Game B is also a losing game in the long run and it has two biased coins: the ``Good'' coin with winning probability $p_g=0.75 - \epsilon $, and the ``Bad'' coin with winning probability $p_b = 0.1- \epsilon$. If the player's capital is a multiple of $3$, the bad coin is used, otherwise the good coin is used. Parrondo paradox refers to the situation that the mixture of A and B game in a sequence leads to winning in the long run. The paradox can be resolved using Markov chain analysis. If the average yield of a game at time $t$ is $\left\langle X \left( t \right) \right\rangle$, then the expected capital gain for game A and B are respectively given by
\begin{eqnarray}
g_A = \left\langle X \left( t+1 \right) \right\rangle - \left\langle X \left( t \right) \right\rangle =2p-1 \label{eqa:1}   \\
g_B = \left\langle X \left( t+1 \right) \right\rangle - \left\langle X \left( t \right) \right\rangle = 2\left(\pi_0 \left(t \right) p_b + \left(1-\pi_0 \left(t \right)\right)p_g \right)-1 \label{eqa:2}
\end{eqnarray}
Note that game A's expected gain is $0$ if $\epsilon=0$, and it is a losing game if $\epsilon>0$. For game B, the $\pi_0 \left(t \right)$ is the probability that the player's capital is multiple of $3$ at time $t$. We can compute the probability vector $\bm{\pi}\left(t \right) = \left( \pi_0 \left(t \right), \pi_1 \left(t \right),  \pi_2 \left(t \right) \right)$, where $\pi_i \left(t \right)$ is the probability that the player's capital is multiple of $3$ plus $i$. To find these probabilities, we consider that transition matrix between different capital states $\left(0,1,2\right)$. For game A and B, the transition matrix that acts on $\bm{\pi} \left(t \right)$ are 
\begin{equation}
\bm{\Pi}_A = 
\begin{pmatrix}
0 & p & 1-p \cr
1-p & 0 & p \cr
p & 1-p & 0
\end{pmatrix},
\bm{\Pi}_B = 
\begin{pmatrix}
0 & p_b & 1-p_b \cr
1-p_g & 0 & p_g \cr
p_g & 1-p_g & 0
\end{pmatrix}
\label{eqa:3}
\end{equation}
so that the evolution equation of the probability vector is $\bm{\pi} \left(t+1 \right)=\bm{\pi} \left(t \right) \bm{\Pi}_G$ where G can be A or B depending on the game played at time $t$. At long time we obtain the stationary distribution state vector $\bm{\pi} \left(t \right)$. In the original Parrondo game, one can show that playing A alone or B alone will lose, but suitable combinations of A and B can lead to positive gain. For a player who can switch his choice of game each time, he can search for a sequence of A and B games to optimize his gain. Indeed, if he knows which game is A, which is B and knows how to count his capital, then he knows how to win maximally. He simply uses the following strategy to make the optimal gain: play game B if his capital is not divisible by $3$, since then his coin is good with a winning probability $p_g>0.5>p$; otherwise play game A as he can avoid the bad coin $p>p_b$. However, if he does not know his capital, or even the game he plays is A or B, then he has a problem in choosing his strategy. Nevertheless, he can still switch the games he will play. For example, he is offered two games called C and D game and he is not told if (C, D) corresponds to (A, B) or (B, A), will he find a way to win? We will provide an answer to this question in this paper.
\par
One possible answer to this question for an ignorant player without memory is to perform random switching of the (C, D) game. Here by random switching we refer to the situation that the player plays game C with probability $\gamma$, or otherwise plays game D. The stochastic combination of game C and game D can be viewed as a new game, which transition matrix is given by $\bm{\Pi} = \gamma \bm{\Pi}_C + \left( 1- \gamma \right) \bm{\Pi}_D$. Now, if he is told that C is A and D is B, then this new game is actually described by the transition matrix $\bm{\Pi} \left( \textnormal{C=A, D=B} \right) = \gamma \bm{\Pi}_A + \left( 1- \gamma \right) \bm{\Pi}_B$. In this way, he can compute the steady state vector and the expected gain. We see that the amount of accessible information has great influence on the long-term return of the player. If the ignorant player learns the above analysis, then he of course can run his numerical experiment assuming one of these scenarios, and compare his gain after some rounds of games, thereby deduce the correct transition matrix that can generate higher capital gain. In another word, the ignorant player, even without help from a friend who knows the mapping of (CD) to (AB) game, can still identify correctly this mapping from his numerical experiments. 
\section{Edward's AB game with one-step memory}
As discussed in the previous section, when the player does not know the identity of the game and cannot count his capital to decide if it is divisible by $3$, he can still do random switching with the switching probability $\gamma$. Of course, if he knows how many games he will be playing, say $N$ games, then he has $2^N$ possible sequences taken from the \{C, D\}. These sequences are deterministic. In principle, he can perform an exhaustive search for small $N$ and use other search algorithms included genetic algorithm for large $N$ sequences~\cite{tang_investigation_2004}. Here we would like to endow the player with finite short-term memory, so that he has some extra information in his decision on the choice of game played next, as opposed to random switching or deterministic switching in a pre-set sequence. This extra information from his memory of the result of the last game enables him to make a ``wiser'' decision, which hopefully yields better performance in the long run. This is the feedback control implemented by the player due to his ability to remember. Note that we assume that the player does not know the current capital and he cannot deduce the current capital state from his finite memory, although he knows the identity of the game. 
\par
\begin{figure}[t]
\centering
\includegraphics[width=0.65\textwidth]{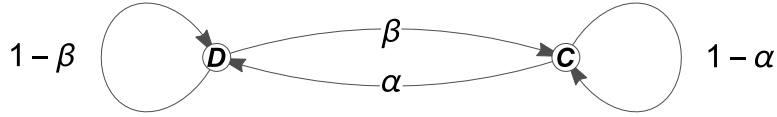}
\caption{At each time step, the player has a probability to switch his choice of the game from A to B or vice versa, conditional on the previous game he played. For example, the player has probability $\alpha$ switching from A to B if his previous game is A; and a probability $\beta$ switching to game A if his previous game is B.}
\label{fig:1}
\end{figure}

We are considering the AB game for which information regarding the software (A/B) of the slot machine (C/D) is accessible to the player. He knows the nature of the game he plays, for example, C is A and D is B. This player is Edward. We now want to find a way for him to optimize his gain. Let's now describe the state of the player with the probability distribution vector $\mathbf{U}\left(t \right) = \left(\pi^A_0\left(t\right),\pi^A_1\left(t\right),\pi^A_2\left(t\right),\pi^B_0\left(t\right),\pi^B_1\left(t\right),\pi^B_2\left(t\right) \right)$. Here A/B means that the player plays the game A/B at time $t$. The subscripts $0,1,2$ denote that the player's capital being a multiple of $3$ plus $0,1,2$ respectively. For example, $\pi^A_0 \left(t \right)$ means that the probability that the player plays game A at time $t$ and his capital is divisible by $3$ so that the remainder is $0$. For one-step memory, we need only to consider the conditional probability $P\left(G_t|G_{t-1} \right)$ for switching. Here the two game sequences played by Edward is denoted by $\left(G_t|G_{t-1} \right)$ which can come in one of the following four scenarios: $\left(G_t|G_{t-1} \right) \in \left\lbrace \left(A_t|A_{t-1}\right),\left(B_t|A_{t-1}\right),\left(A_t|B_{t-1}\right),\left(B_t|B_{t-1}\right) \right\rbrace$. We introduce here two parameters $\alpha,\beta$, as the probability of switching as illustrated in \fref{fig:1}.
\par
\begin{figure}
\centering
\includegraphics[width=0.55\textwidth]{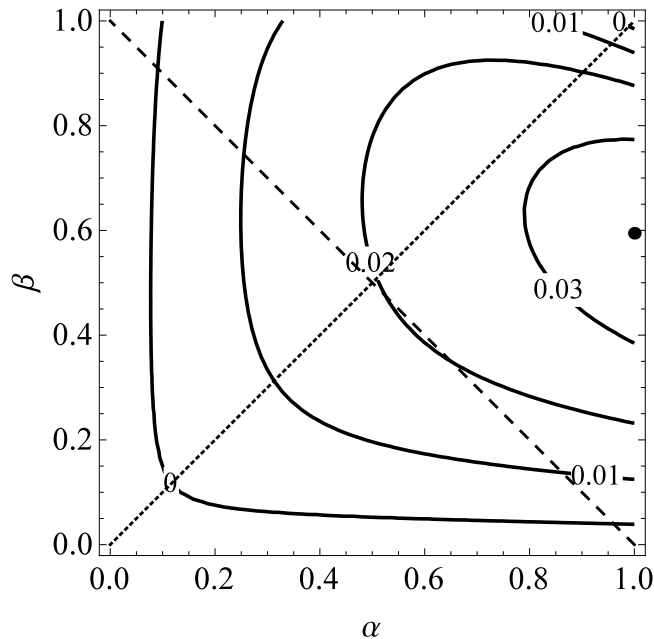}
\caption{The contour plot of the expected capital gain for the Parrondo game for player with one step memory in the game sequence with switching probabilities $\alpha$ and $\beta$. Here $p=0.5-\epsilon, p_b=0.1-\epsilon, p_g=0.75-\epsilon, \textnormal{ with } \epsilon=0.003$. For this set of winning probabilities, the maximum long-term gain is achieved by the switching probabilities $\left(\alpha=1, \beta=0.5907 \right)$, yielding a gain of $0.0355$. The search space for the optimal expected gain is defined by the dotted line for CD game with $\alpha=\beta$. On the other hand, the search space for the optimal expected gain in random switching is defined by the dash line, with $\alpha=1-\beta, \beta = \gamma$.}
\label{fig:2}
\end{figure}

Under this hypothesis of conditional probability for one-step memory and using the switching scheme in \fref{fig:1}, we can represent the game for player with one-step memory by a $6 \times 6$ transition matrix $\bm{\Pi}_s$. The transition elements can be written as $P\left(G_t | G_{t-1} \right) P_{win} \left(G_t \right)$ if the game at $t-1$ is winning, and $P\left(G_t | G_{t-1} \right) \left(1-P_{win} \left(G_t \right) \right)$ if the previous game is losing. Since the probability of switching $\left(\alpha, \beta \right)$ is independent from the probability of winning, we can write the transition matrix element as a product of two probabilities. The transition matrix $\bm{\Pi}_s$ is given by 
\begin{equation}
\medmuskip = -3mu
\hspace{-27mm}
\def\quad{\hskip1ex\relax}
\bm{\Pi}_s=
\begin{pmatrix}
0 & \left(1-\alpha \right)p & \left(1-\alpha\right)\left(1-p\right) & 0 & \alpha p & \alpha \left(1-p\right) \cr
\left(1-\alpha\right)\left(1-p\right) & 0 & \left(1-\alpha\right) p & \alpha \left(1-p\right) & 0 &\alpha p \cr
\left(1-\alpha \right)p & \left(1-\alpha\right)\left(1-p\right) & 0  & \alpha p & \alpha \left(1-p\right) & 0 \cr
0 & \beta p_b & \beta \left(1-p_b\right) & 0 & \left(1-\beta \right) p_b & \left(1-\beta \right)\left(1-p_b\right) \cr
\beta \left( 1-p_g \right) & 0 & \beta p_g & \left(1-\beta \right) \left(1-p_g \right) & 0 &\left(1-\beta\right) p_g \cr
\beta p_g & \beta \left(1-p_g \right) & 0 & \left(1-\beta \right)p_g & \left(1-\beta\right)\left(1-p_g \right) & 0
\end{pmatrix}
\label{eqa:4}
\end{equation}
We show the expected capital gain for particular set of winning probabilities in \fref{fig:2}.
\par
The search on the two dimensional parameter space for maximum gain, treating the two switching probability for player with one-step memory as independent parameters, shows that the highest expected capital gain is at the rightmost region of \fref{fig:2}, denoted by the dot which corresponds to the probability of switching $\alpha=1, \beta=0.5907$. 
\par
Now let's first recall in the last section we have discussed random switching by the ignorant player. We first derive the relation of random switching and general switching with one-step memory by considering the joint probability of two consecutive game events. The probability of the event that the player plays game A at time $t-1$ and plays B at time $t$ is given by the Bayes expression with the conditional probability defined in \fref{fig:1} given by $P\left(B_t|A_{t-1} \right)= \alpha$ so that 
\begin{equation}
P\left(B_t \cap A_{t-1} \right) = P\left(B_t|A_{t-1} \right)P\left(A_{t-1} \right)=\alpha P \left(A_{t-1} \right)
\label{eqa:5}
\end{equation}
For random switching, $P\left(B_t \cap A_{t-1} \right) = P\left(B_t \right)P\left(A_{t-1} \right)$ because the event $B_t$ and $A_{t-1}$ are independent, which is $P\left(B_t| A_{t-1} \right)=P\left(B_t \right) = 1-\gamma$, so that
\begin{equation}
P\left(B_t \cap A_{t-1} \right) = P\left(B_t \right) P \left(A_{t-1} \right)=\left(1-\gamma \right)P \left(A_{t-1} \right)
\label{eqa:6}
\end{equation} 
Combine \eref{eqa:5} and \eref{eqa:6} we get $\alpha=1-\gamma$. Similarly, consider the probability $P\left(A_t|B_{t-1} \right)$, we can get $\beta = \gamma$. Thus, random switching satisfies the constraint $\alpha=1-\beta$, which is a special case in the two-dimensional parameter space for the switching with one-step memory. Therefore, the optimal random switching probability for Parrondo game is obtained along the dashed line $\alpha=1-\beta$ in \fref{fig:2}. Note that the random switching strategy cannot yield a higher gain than the optimal gain achieved by player with one-step memory, since the random switching algorithm only \textit{searches in a subset of the entire parameter space} in the $\left(\alpha, \beta \right)$ plane. In fact, the overall optimal switching parameters for the Parrondo game with one-step memory must achieve at least equal or higher capital gain for the random switching case. In \fref{fig:2}, we can find that the maximum gain is achieved with $\left(\alpha=1, \beta=0.5907 \right)$, showing hat higher gain is achieved if the player has one-step memory, which is expected.  In the next section, when we consider CD game where the player has no knowledge of the mapping of (C, D) to (A, B), then our analysis of the gain in \fref{fig:2} corresponds to the search along the subspace $\alpha=\beta$, which is represented the dotted line inside the figure.
\section{Frank's CD game with one-step memory}
Let us now consider the strategy recommended to Frank, who plays on the two slot machines C (for ``Charles'') and D (for ``David'') not knowing the mapping of the software installed, so that CD may correspond to AB or BA. This lack of information on the identity of the software installed in the slot machines implies that Frank must use a different strategy of game switching. Similar to Edward, Frank also cannot keep track of his capital, though he still remembers the slot machine he used in the last game. Now, due to his ignorance of the mapping between CD and AB, he has no reason to switch from D to C with a probability $\beta$ that is different from $\alpha$. Therefore, logically Frank should use the same probability of switching and his solution space for the search of maximum gain for the CD game should be the subset of the two-dimensional parameter space defined by the dotted line $\alpha=\beta$ in \fref{fig:2}. By setting $\alpha=\beta$ into the transition matrix \eref{eqa:4}, when $\alpha \rightarrow 0$ and $\alpha \rightarrow 1$, the corresponding transition matrices $\mathbf{X}$ and $\mathbf{Y}$ are 
\begin{equation}
\mathbf{X} = 
\begin{pmatrix}
\mathbf{0} & \mathbf{\Pi}_A \cr
\mathbf{\Pi}_B & \mathbf{0} 
\end{pmatrix},
\quad
\mathbf{Y} = 
\begin{pmatrix}
\mathbf{\Pi}_A & \mathbf{0} \cr
\mathbf{0} & \mathbf{\Pi}_B 
\end{pmatrix}
\label{eqa:7}
\end{equation}
where $\mathbf{0}$ is a $3 \times 3$ zero matrix. The CD Parrondo game with one-step memory has a transition matrix for switching defined by the linear combination of game X and Y.
\begin{equation}
\mathbf{\Pi}_s = \alpha \mathbf{X} + \left(1-\alpha\right) \mathbf{Y} = 
\begin{pmatrix}
\left(1-\alpha\right) \mathbf{\Pi_A} & \alpha \mathbf{\Pi_A} \\
\alpha \mathbf{\Pi_B} & \left(1-\alpha\right) \mathbf{\Pi_B} 
\end{pmatrix}
\label{eqa:8}
\end{equation}
This form of transition matrix has a simple interpretation. Game X is a game in which the player plays game A and game B alternatively, and game Y is a game in which the player always follows the previous game. If the player plays game X alone, the game sequence is CDCD...; if the player plays game Y alone, the game sequence is either CCC... or DDD..., depending on the starting game of Y. With this separation of the general transition matrix into a combination of X and Y, we can interpret the Frank`s CD game with one-step memory as the random switching of game X and game Y. From this perspective, CD game can be interpreted as another Parrondo's paradox, allowing the combination of two losing games X and Y that leads to a win in the long run.
\begin{figure}[t]
\centering
\includegraphics[width=0.6\textwidth]{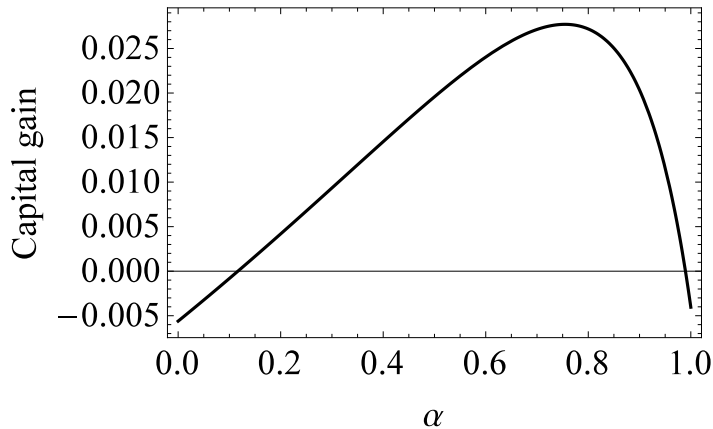}
\caption{The expected capital gain for CD Parrondo game with either CD=AB or BA. The parameters for the CD game is same as the original Parrondo game with $\left(p=0.5-\epsilon, p_g=0.75-\epsilon, p_b=0.1-\epsilon \right)$ and $\epsilon=0.003$. The capital gain is concave leading to positive gain for some values of $\alpha$.} 
\label{fig:3}
\end{figure}
\par
\begin{figure}
\centering
\includegraphics[width=0.55\textwidth]{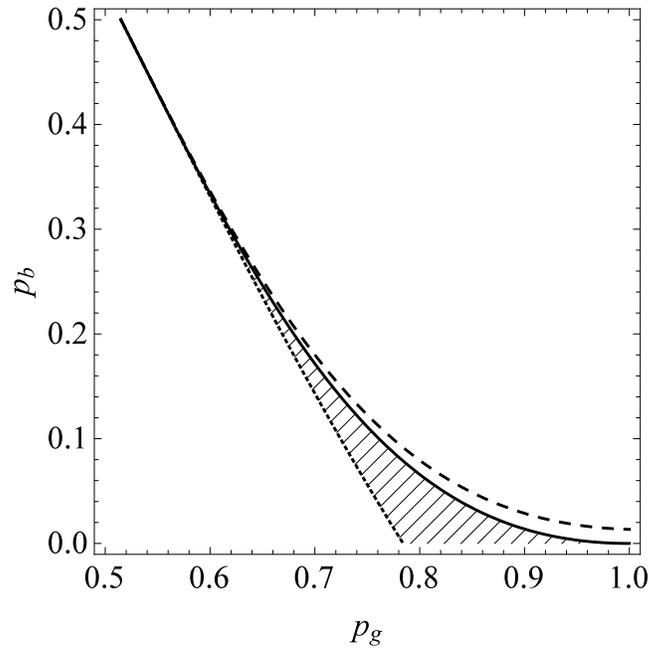}
\caption{The Win/Loss region of Parrondo game with $p=0.49$. The three lines corresponds to the fair game boundaries for three different values of $\alpha$. (solid line: $\alpha=1$; dash line: $\alpha=0$; dotted line: $\alpha=0.5$). The fair game boundary is defined by the condition that the region at the right hand side of the fair game boundary corresponds to the winning region of parameter space $\left(p=0.49, p_g, p_b \right)$ when the CD game is played with the associated switching parameter $\alpha$, while the game is losing in the long run for parameters in the region left of the boundary. In this figure we see that the winning region for $\alpha=0.5$ is larger than the winning region for $\alpha=0$ as well as the winning region for $\alpha=1$. The shaded area corresponds to the region of the parameter space of CD game such that the player with switching probability $\alpha=0.5$ is winning but the game with $\alpha=0$ or $1$ is losing. This shaded area corresponds to the region of the parameter space of CD game exhibiting strong Parrondo effect.}
\label{fig:4}
\end{figure}

To analyze the performance along the line $\alpha=\beta$, we can calculate the long term expected gain and the result is shown in \fref{fig:3}. We can see that the function of expected capital gain is concave along the axis $\alpha$. Although the two extreme case $\left(\alpha=0 \textnormal{ and } \alpha=1 \right)$ are both losing (CDCD...is losing, CCCC... or DDDD... are also losing), we can still make a positive gain by mixing game X and game Y using some intermediate $\alpha$. This is a generalized Parrondo effect for the player with memory, though without the knowledge of the CD to AB mapping. We can now use this transition matrix and work out the fair game boundary for a given set of parameters $\left(p,p_g,p_b \right)$. Since $p$ is the coin parameter for A game, we can assume it to be $0.49$, which means that the Casino takes a $2\%$ commission $\left(g=-2\epsilon=-0.02 \right)$. The effective two-dimensional parameter space is illustrated in \fref{fig:4} with x-axis defined by $x=p_g>0.5$ for the winning probability for the good coin and the y-axis defined by $y=p_b<0.5$ for the bad coin. This parameter space can be divided into two regions by the fair game boundary (\fref{fig:4}). Here, the fair game boundary is defined by the condition that the region at the right hand side of the fair game boundary (large $p_g$ for fixed $p_b$) corresponds to the winning region of parameter space $\left(p=0.49, p_g, p_b \right)$  when the CD game is played with the associated switching parameter $\alpha$. Similarly, the game is losing in the long run for parameters in the region left of the boundary (smaller $p_g$ for fixed $p_b$.) Each fair game boundary is also parameterized by a given switching probability $\alpha$.
\par
In \fref{fig:4}, we have drawn three fair game boundaries for three cases of $\alpha=0,0.5 \textnormal{ and } 1$. The three winning regions are different. Noticeable is the fair game boundary for $\alpha=0.5$ lies on the left of both the fair game boundary for $\alpha=0$ and for $\alpha=1$. This implies that the winning region for the $\alpha=0.5$ case is bigger than either $\alpha=0$ and $\alpha=1$. The shaded area corresponds to the region of the parameter space of CD game such that the player with switching probability $\alpha=0.5$ is winning but the game with $\alpha=0$ or $1$ is losing. Since $\alpha=0$ corresponds to no switching (game Y) while $\alpha=1$ corresponds to always switching (game X), so that with appropriate switching such as $\alpha=0.5$, we can combine two losing CD game (X and Y are both losing) into a winning one. The above analysis for the fair game boundary in \fref{fig:4} allows one to conclude that there exists region in the parameter space $\left(p=0.49,p_b, p_g\right)$ where a CD game player can win using $\alpha=0.5$, but he can lose using some other values of $\alpha$. This observation corresponds to the definition of strong Parrondo effect~\cite{wu_extended_2014} where the combination of two losing game leads to a winning one.
\section{Optimal Switching Probability for the CD Parrondo game}
Since the CD Parrondo game has strong Parrondo effect, it is natural to ask if there is an optimal switching probability for the CD game when the underlying software is the original ``AB$3$'' Parrondo game with A being a fair coin tossing and B with a good and a bad coin coupled with the state of capital modulo $3$. The answer turns out to be very interesting, as the result is universal in the sense that the optimal switching probability for the CD game is exactly $\alpha=3/4$ and is independent of $\left(p_g, p_b \right)$ as long as $p=0.5$ and $\left(p_g>0.5, p_b<0.5 \right)$. Unfortunately, this universal switching probability only works for AB3 game, and the optimal switching probability for other AB$M$ game (modulo $M$ for $M$ different from $3$) depends on the parameters $\left(p_g>0.5, p_b<0.5\right)$.
\par
To derive the optimal $\alpha$ for ``AB$3$'' Parrondo game, we first need to solve for the state vector $\mathbf{U}$ in the following system of equations, 
\begin{equation}
\left(\bm{\Pi}_s^T - \mathbf{I}\right) \mathbf{U}^T = \mathbf{0} \textnormal{ and } \sum_{i=1}^6 \pi_i\left(t \right)=1
\label{eqa:9}
\end{equation}
where $\mathbf{U} = \left(\pi^A_0,\pi^A_1,\pi^A_2,\pi^B_0,\pi^B_1,\pi^B_2 \right)$ is the  vector of stationary probabilities. 
The optimal value of $\alpha$ which maximizes the expected return for the AB3 game is shown in the appendix to be 
\begin{equation}
\alpha^* = \frac{\left(1-p+p^2 \right)\left(2p_g-1 \right)}{4p^2p_g-2p^2-4pp_g+3p_g+3p-2}
\label{eqa:10}
\end{equation}
Note that when $p=0.5$, meaning that game A is fair, then the optimal value $\alpha$ becomes a constant value of $3/4$. Therefore switching between C and D with probability $3/4$ is an optimal way to play the AB3 game if the player cannot distinguish the game A and game B. The corresponding solution for the probability distribution vector $\mathbf{U}\left(t \right)$ for $t \rightarrow \infty$ at $\alpha=\alpha^*=3/4$ is
\begin{equation}
\mathbf{U} \left(t \right)_{a^*} = \left( \frac{1}{6}, \frac{1}{18} \left(3+2p_b-2p_g \right), \frac{1}{18} \left(3-2p_b+2p_g \right), \frac{1}{6}, \frac{1}{6}, \frac{1}{6} \right)
\label{eqa:11}
\end{equation}
Since the last three terms correspond to the capital distribution when playing game B, we see that the effect of optimal switching is to redistribute the capital state at equal probability. In another word, the optimal switching scheme achieves equal probability of the occupation of the state with capital remainder being $0,1$ or $2$ when playing B in the long run.
\section{Discussion}
In this paper we consider two kinds of players (Edward and Frank) entering a casino with two slot machines called Charles (C) and called David (D). Both are poor in mathematics and cannot keep track of the capital they have, but can remember which slot machine they used in the last game, (i.e., both are one-step memory players. Edward has more information about the slot machines for he knows the software (A, B) installed in them, but Frank does not even have this information. We solve for the strategies that Edward and Frank should use with the information given to optimize their capital gain. 
\par
For the particular sets of winning probabilities used in the original Parrondo game (AB3), ($ p=0.5-\epsilon, p_b=0.1-\epsilon, p_g=0.75 -\epsilon $ with $\epsilon=0.003$ and modulo $3$ for the state of capital), the optimal frequency for switching for Edward is $\left(\alpha=1, \beta=0.5907 \right)$. This implies that the one-step memory player will always switch his game to B right after a game A. This result agrees with the optimal periodic sequence ABABB discovered in \cite{dinis_optimal_2008} and \cite{tang_investigation_2004}. 
\par
Our switching scheme can be generalized to player with n-step memory by setting the conditional probability $P\left(G_t | G_{t-1}^{\left(n\right)} \right)$ for different history conditions, where $G_{t-1}G_{t-2} \dots G_{t-n}$ is the game history with length $n$. For example, in two-step memory, there are four history combinations $\left\lbrace A_{t-1} A_{t-2}, A_{t-1}B_{t-2}, B_{t-1}A_{t-2}, B_{t-1} B_{t-2} \right\rbrace$, the matrix dimension is $4 \times 3 = 12$ (here the $4$ refers to the four history combinations and the $3$ refers to the three capital states: $0,1,2$). In general, the dimension for a transition matrix with n-steps memory is $\left(3 \cdot 2^n \right) $. 
\par
In our analysis, we have focused on the game B with modules $M=3$, and investigate the Parrondo effect in the parameters space $\left(p_g>0.5, p_b<0.5 \right)$ and switching probabilities $\alpha, \beta$. By numerical calculation, we can find the Parrondo effect also exists for larger $M$ by choosing suitable sets of parameters. Here we need to consider the even $M$ and odd $M$ cases separately since the results are very different.
\par
For odd $M$, there exists Parrondo effect. We consider the special case when $p=0.5$ so that A game corresponds to diffusion and  $\mathbf{\Pi_A}$ in \eref{eqa:3} becomes a unbiased random walk process. When $\alpha \rightarrow 1$, game A will not alter $\mathbf{\pi_0}$ in the equilibrium distribution in game B and its payoff, so that  the expected gain for BBB\dots and ABAB\dots will be the same.  This means that in the long run, the case for $\alpha \rightarrow 0$ which corresponds to BBB\dots will yield the same expected return as the case when $\alpha \rightarrow 1$ which corresponds ABAB\dots sequence. Next, because the capital gain for CD game is in general a non-linear function of $\alpha$, we expect that there exists some intermediate $0<\alpha<1$ leading to a higher or lower payoff than that when playing B games alone, depending on the winning parameters. If the payoff is higher, then Parrondo effect exists \cite{wu_extended_2014}.
\par
\begin{figure}
\centering
\includegraphics[width=0.8\textwidth]{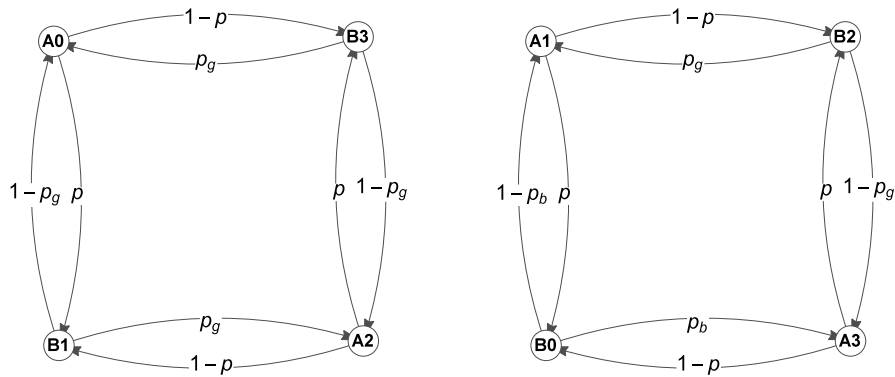} 
\caption{The irreducible Markov chains for different initial conditions. We first discuss the figure on the left. If the initial game is game A with even initial capital, we start with $A0$ or $A2$, the game will be trapped into the chain on the left. Similarly, if the initial game is B for the odd initial capital, we start with $B1$ or $B3$, the game will be trapped also in the figure on the left. Otherwise, the game sequence will follow the figure on the right. }
\label{fig:5}
\end{figure}

However, for even $M$, the existence of Parrondo effect depends on the initial capital, as the Markov chain for the even $M$ case is reducible. For example, when $M=4$ and $p=0.5, p_g>0.5, p_b<0.5$ and when the first game is A and capital is even at $t=0$,  
the highest capital gain is achieved with the extreme value of switching probability $\alpha \rightarrow 1$. The corresponding game sequence is ABAB\dots . The reason is with initial capital being even, the first game chosen by the player to maximize his probability of winning will be A as $p=0.5> p_b$. Then, for the next game, his capital can only be $\pm 1$, so that he will play the B game since $p_g>p=0.5$. This will lead to the ABAB\dots as the maximum average return sequence. In the Markov chain terminology, for different initial condition, it will be trapped in one of the two different irreducible Markov chains which is shown in \fref{fig:5}.
\par
In \fref{fig:5}, we foresee four situations: 
\begin{enumerate}
\item the initial capital is even and start with A; 
\item the initial capital is odd and start with B; 
\item the initial capital is even and start with B; 
\item the initial capital is odd and start with A; 
\end{enumerate}
Case (i) and (ii) corresponds to the left Markov Chain shown in \fref{fig:5}, while case (iii) and (iv) corresponds to the right Markov Chain. 
For case (i), the state of capital is even, the state $A0$ means the player is playing game A when the capital modulo is $0$ . Thus, when the initial capital is even, the game sequence starting with game A will be trapped in left Markov chain. In this case, the player avoids the use of bad coin $p_b$, therefore he can win more than any sequence resulted from $\alpha<1$ . Now, when the switching probability $\alpha=1$, the sequence is a deterministic sequence ABAB\dots and there is no Parrondo effect. The same argument applies to the case when the initial capital is odd and the game sequence starts with B, and the Markov Chain is also the left one in \fref{fig:5}. 
On the other hand, if we start with game B and the initial capital is even, then there can be Parrondo effect. These arguments provide heuristic justification for our numerical calculation for the even $M$ cases. 
\par 
The relation between the Brownian ratchet and the Parrondo Game has been discussed \cite{harmer_parrondos_1999, parrondo_brownian_2004, amengual_discretetime_2004, toral_parrondos_2003,allison_physical_2002}. They have identified a correspondence of a set of quantities between games and flashing ratchets and the similarity is both mathematically pleasing and physically reasonable. One interesting analogy is the sequence of switching potential in flashing ratchets, and the change of games by the player \cite{harmer_parrondos_1999}. The A game corresponds to the potential being in the OFF state, while the B game corresponds to the ON state of the asymmetric potential in the ratchet problem. Indeed, the switching of the games with probability $\alpha, \beta$ can be viewed as the discrete version of the fluctuating potential in flashing ratchet modelled by the continuous Markov process \cite{bao_langevin_1998,abad_brownian_1998}.
\begin{equation}
\frac{d}{dt}
\begin{pmatrix}
P\left(\textnormal{on},t\right) \cr
P\left(\textnormal{off}, t\right) 
\end{pmatrix}
=
\begin{pmatrix}
-\lambda_\textnormal{on} &  \lambda_\textnormal{on} \cr
\lambda_\textnormal{off} & -\lambda_\textnormal{off}
\end{pmatrix}
\begin{pmatrix}
P\left(\textnormal{on},t\right) \cr
P\left(\textnormal{off}, t\right) 
\end{pmatrix}
\end{equation}
where $P\left(\textnormal{on},t\right)$ is the probability that the potential is on at time $t$ and $\lambda_{\textnormal{on}}, \lambda_{\textnormal{off}}$ are the flipping rate. By performing Langevin simulation on the ratchets, we can see that an optimal $\lambda$ which maximizes the particle current does exist. The optimal value of $\lambda$ also depends on the shape of the potentials. Therefore, the conditional switching with memory in the game sequence is similar to the ratchets with different $\lambda_{\textnormal{on}}$ and $ \lambda_{\textnormal{off}}$.
\ack K. Szeto acknowledges the support of grant FSGRF13SC25 and FSGRF14SC28 
\section*{References}
\bibliography{UPON}
\section*{Appendix}
To derive \eref{eqa:10}, we solve the system of equations in \eref{eqa:9}
\begin{equation}
\medmuskip = -3.5mu
\hspace{-28mm}
\def\quad{\hskip1ex\relax}
\begin{pmatrix}
-1 & \left(\alpha-1 \right)\left( p-1\right) & \left(1-\alpha\right)p & 0 & \alpha \left(1-p_g\right) & \alpha p_g \cr
\left(1-\alpha\right)p & -1 & \left(\alpha-1\right) \left(p-1 \right) & \alpha p_b & 0 & \alpha \left(1-p_g\right)  \cr
\left(\alpha-1 \right)\left(p-1\right) & \left(1-\alpha\right)p & -1  & \alpha \left(1-p_b \right) & \alpha p_g & 0 \cr
0 & \alpha \left(1-p \right) & \alpha p & -1 & \left(\alpha-1 \right) \left( p_g-1 \right) & \left(1-\alpha \right)p_g \cr
\alpha p & 0 & \alpha \left(1-p\right) & \left(1-\alpha \right) p_b & -1 & \left(\alpha-1\right) \left( p_g-1 \right) \cr
1 & 1 & 1 & 1 & 1 & 1
\end{pmatrix}
\mathbf{U^T} =
\begin{pmatrix}
0 \cr 0 \cr 0 \cr 0 \cr 0 \cr 1
\end{pmatrix}
\end{equation}
where 
$\mathbf{U} = \left(\pi^A_0,\pi^A_1,\pi^A_2,\pi^B_0,\pi^B_1,\pi^B_2 \right)$.
After some lengthy calculation we get for $\pi^B_0$ the following long equation,
\begin{equation}
\pi^B_0 = \frac{F_2 \alpha^2 + F_2 \alpha + F_0}{G_2 \alpha^2 + G_1 \alpha  + G_0} 
\end{equation}
where the coefficients are 
\begin{equation}
\hspace{-20mm}
\begin{aligned}
F_2 &= 12 p^2 p_g^2-12 p^2 p_g-12 p p_g^2+14 p p_g+7 p_g^2-8 p_g+7 p^2-8 p+4 \\
F_1 &= -12 p^2 p_g^2+12 p^2 p_g+12 p p_g^2-12 p p_g-9 p_g^2+9 p_g-9 p^2+9 p-6 \\
F_0 &= 3 p^2 p_g^2-3 p^2 p_g-3 p p_g^2+3 p p_g+3 p_g^2-3 p_g+3 p^2-3 p+3 \\
G_2 &= 48 p^2 p_b p_g-48 p p_b p_g+28 p_b p_g-24 p^2 p_b+28 p p_b-16 p_b+24 p^2 p_g^2
\\ &\quad-48 p^2 p_g-24 p p_g^2+56 p p_g+14 p_g^2-32 p_g+42 p^2-48 p+24\\
G_1 &= -48 p^2 p_b p_g+48 p p_b p_g-36 p_b p_g+24 p^2 p_b-24 p p_b+18 p_b-24 p^2 p_g^2
\\ &\quad +48 p^2 p_g+24 p p_g^2-48 p p_g-18 p_g^2+36 p_g-54 p^2+54 p-36\\
G_0 &= 12 p^2 p_b p_g-12 p p_b p_g+12 p_b p_g-6 p^2 p_b+6 p p_b-6 p_b+6 p^2 p_g^2-12 p^2p_g
\\ &\quad -6 p p_g^2+12 p p_g+6 p_g^2-12 p_g+18 p^2-18 p+18
\end{aligned}
\label{eqa:15}
\end{equation}
In order to get the optimal value $\alpha$, one should minimize $\pi_0^B$ for the following reason: the $\alpha$ that maximizes the probability $\pi_1^B $ and $\pi_2^B $ will maximize the capital gain since they use the good coin and game A is fair. Now, in the CD Parrondo game, the proportion of games played being game A is the same as being B. This observation implies that
\begin{equation}
\pi_0^B +\pi_1^B +\pi_2^B  = 1/2 = \pi_0^A  + \pi_1^A  + \pi_2^A 
\end{equation}
Now, we are interested only in state vector at long time. In our computation of the expected gain in the long run, we can use $\pi_1^B+ \pi_2^B = 1/2 -\pi_0^B $ to replace maximizing $\pi_1^B  \textnormal{ and } \pi_2^B $ by minimizing $\pi_0^B$. Based on the expression of the capital gain which is a function of $\left(\pi^A_0,\pi^A_1,\pi^A_2,\pi^B_0,\pi^B_1,\pi^B_2 \right)$ and $\alpha$, we can see that the maximization of gain for the CD game is equivalent to the minimization of $\pi_0^B $. Once this observation is made, we can simply take the derivative of the $\pi_0^B$ with respect to $\alpha$ to get the optimal value $\alpha=\alpha^*$. This symmetry argument allows us to get the optimal $\alpha^*$ that maximizes the expected gain for the CD game, by setting the derivative of the gain with respect to $\pi_0^B$ to zero. Then we can get
\begin{equation}
\alpha = \frac{\left(1-p+p^2 \right)\left(2p_g-1 \right)}{4p^2p_g-2p^2-4pp_g+3p_g+3p-2}
\end{equation}
which is the optimal value α that makes $\pi_0^B$ stationary. For $p=0.5$, we get the unique solution 3/4. Substitute $\alpha=\alpha*=3/4$ into the system of equations, we can solve the remaining terms in $\mathbf{U}$, and finally we can get \eref{eqa:11}.
We can see that when $p=0.5$, \eref{eqa:15} can be simplified as 
\begin{equation}
\begin{aligned}
F_2 &= 4 p_g^2-4 p_g+\frac{7}{4} \\
F_1 &= -6 p_g^2+6 p_g-\frac{15}{4} \\
F_0 &= \frac{9 p_g^2}{4}-\frac{9 p_g}{4}+\frac{9}{4} \\
G_2 &= 16 p_b p_g-8 p_b+8 p_g^2-16 p_g+\frac{21}{2} \\
G_1 &= -24 p_b p_g+12 p_b-12 p_g^2+24 p_g-\frac{45}{2} \\
G_0 &= 9 p_b p_g-\frac{9 p_b}{2}+\frac{9 p_g^2}{2}-9 p_g+\frac{27}{2}
\end{aligned}
\end{equation}
\end{document}